\documentclass[a4paper,12pt]{article}
\usepackage[pctex32]{graphics}
\usepackage{amssymb,amsmath}
\usepackage{amsmath,amssymb}
\usepackage{latexsym}
\usepackage{epsfig}
\usepackage[english]{babel}

\newcommand{\be}{\begin{equation}}
\newcommand{\ee}{\end{equation}}
\newcommand{\ba}{\begin{eqnarray}}
\newcommand{\ea}{\end{eqnarray}}

\begin{document}

\begin{titlepage}

\vspace{5mm}

\begin{center}

{\Large \bf Renormalizability and Newtonian potential  \\ in  scale-invariant  gravity}

\vskip .6cm

\centerline{\large
 Yun Soo Myung$^{a}$}

\vskip .6cm

{Institute of Basic Science and Department of Computer Simulation,
\\Inje University, Gimhae 50834, Korea \\}

\end{center}

\begin{center}
\underline{Abstract}
\end{center}
There is a conjecture that  renormalizable higher-derivative gravity has  a finite classical potential at the origin.
In this work we show clearly that the scale-invariant gravity (SIG)  satisfies the conjecture.
This gravity produces the better-behaved $1/k^4$ UV behavior as needed for renormalizability.
It turns out that the SIG has the linear classical potential of $V\propto r$ and it is a UV complete theory.
 \vskip .6cm

\vskip 0.8cm

\vspace{15pt} \baselineskip=18pt

\noindent $^a$ysmyung@inje.ac.kr \\

\thispagestyle{empty}
\end{titlepage}

\newpage
%%%%%%%%%%%%%%%%%%%%%%%%%%%%%%%%%%%%%%%%%%%%%%%%%%%%%%%%%%%%%%%%%%
\section{Introduction}
There was a conjecture that renormalizable higher-derivative gravity has  a finite Newtonian potential at the  origin~\cite{Modesto:2014eta}.
This relation was first mentioned  in Stelle's seminal work~\cite{Stelle:1976gc} which showed that the fourth-derivative gravity (FDG) is renormalizable and  has a finite potential at origin.
However, this gravity  belongs to a nonunitary theory because it has a massive pole with negative residue which could be interpreted  either as a state of negative norm or a state of negative energy.
 In this case, a massive ghost tensor  and a massive healthy scalar  contribute in such a manner to cancel out the Newtonian singularity originated from a massless tensor.

Recently, it was  conjectured that the reverse of the above statement is not true~\cite{Giacchini:2016xns}.
It implies  that the finiteness of the Newtonian potential at the origin is a necessary but not sufficient condition for the renormalizability of the model.
The models considered in~\cite{Giacchini:2016xns,Accioly:2017rft} are different versions of  sixth-derivative
gravities. One is a fully six-derivative gravity which satisfies the above conjecture, whereas the other is a half six-derivative gravity which does not satisfy the
reverse of the statement. That is, the latter is the case that even though the potential is finite at $r=0$, it is known to be  non-renormalizable by power counting.
More recently, it was probed that the necessary condition for (super-)renormalizability implies the sufficient condition for the cancellation of the Newtonian singularities in $D\ge 4$-dimensions
~\cite{Accioly:2017xmm}.

In this work, we will show clearly that the scale-invariant gravity (SIG)   satisfies the above  conjecture.
This gravity  is considered as  a very suitable model  for  exploring the connection between  classical  potential  and renormalizability of the theory.
We note that this  theory  has  the better-behaved $1/k^4$ tensor and scalar propagators  as needed for renormalizability.
It turns out that the SIG  provides the linear classical potential of $V\propto r$ and it is a UV complete theory.

\section{Scale-invariant gravity(SIG)}
 A classically scale-invariant   gravity in four dimensions is defined by
 \begin{equation} \label{action1}
S_{\rm SIG}=\int d^4x \sqrt{-g}\Big[-\frac{1}{a}\Big(\frac{1}{2}C^2_{\mu\nu\rho\sigma}+\frac{w}{3}R^2\Big)+c GB\Big]
 \end{equation}
where $C_{\mu\nu\rho\sigma}$ is the Weyl tensor and $R$ is the Ricci scalar. Also $GB=R^2_{\mu\nu\rho\sigma}-4R^2_{\mu\nu}+R^2$ is the Gauss-Bonnet topological term  expressed by the divergence of a current $B^\mu$ as $GB=\nabla_\mu B^\mu$. Hence it makes no contribution to the equations of motion and we may ignore it when formulating the Feynman rules for the quantum theory.
In other words, $GB$ does not contribute to the classical potential. One notes that the effect of this term  is not relevant for one-loop renormalization, but it  may play a role in renormalization group equations.
  The three coupling  constants of  $a$ (Weyl coupling), $w$ (Ricci coupling), and $c$ (Gauss-Bonnet coupling) are dimensionless and hence it is easy to derive their renormalization group equations.

   The action (\ref{action1}) without $GB$ was first introduced as a gravity  model for admitting linearized solutions with negative energy but all exact solutions possess zero energy~\cite{Boulware:1983td}.
On later, this was considered continuously as a prototype for renormalization and renormalization group analyses~\cite{Fradkin:1981iu,David:1984uv,Avramidi:1985ki,deBerredoPeixoto:2004if,Tomboulis:2015esa}.
Hence, the SIG is considered as  a very suitable model  for  exploring the close connection between  classical  potential  and renormalizability of the theory.

On the other hand, we note that if one includes the Einstein-Hilbert term $2R/\kappa^2$ with $\kappa^2=4\kappa_4(\kappa_4=8\pi G)$, the corresponding action is not scale-invariant and
 was  employed to prove the renormalizability of the FDG~\cite{Stelle:1976gc}. Furthermore, ignoring $R^2$ leads to the conformal (Weyl)
 gravity with the Gauss-Bonnet term. This model was first considered in~\cite{Fradkin:1981iu}, whereas  on later two of {\cite{Antoniadis:1992xu,deBerredoPeixoto:2003pj} have checked and corrected~\cite{Fradkin:1981iu}.

At this stage, we introduce the well-known relation
\begin{equation} \label{rel1}
C^2_{\mu\nu\rho\sigma}=2W+GB,~ W=\Big(R^2_{\mu\nu}-\frac{1}{3}R^2\Big)
\end{equation}
which implies  $ C^2=2W$ up to total divergence.
Using (\ref{rel1}), (\ref{action1}) can be written as
 \begin{equation} \label{action2}
\tilde{S}_{\rm SIG}=\int d^4x \sqrt{-g}\Big[-\frac{1}{a}R^2_{\mu\nu}+\frac{1}{3a}(1-w)R^2+(c-\frac{1}{2a}) GB\Big],
 \end{equation}
 which is useful to derive the equation of motion and the  propagator.
In order to find the classical potential, one has first  to obtain the propagator. For this purpose,
we expand the metric tensor  $g_{\mu\nu}= \eta_{\mu\nu}+h_{\mu\nu}$  around the Minkowski metric $\eta_{\mu\nu}=(+---)$.
Bilinearizing the Lagrangian of  Eq. (\ref{action2})  together with  imposing the de Donder gauge,
one obtains ${\cal L}^{\rm bil}_{\rm SIG}=h^{\mu\nu}{\cal O}_{\mu\nu,\alpha\beta}h^{\alpha\beta}/2$~\cite{Accioly:2017rft,Stelle:2017bdu}
with
\begin{equation}
{\cal O}_{\mu\nu,\alpha\beta}=-\frac{1}{2a} \square^2 P^{(2)}_{\mu\nu,\alpha\beta}-\frac{2w}{a}\square^2P^{(0-s)}_{\mu\nu,\alpha\beta}+\cdots.
\end{equation}
Inverting ${\cal O}$,  we obtain the propagator for the SIG in momentum space
\begin{equation} \label{propto1}
{\cal D}^{\rm SIG}_{\mu\nu,\alpha\beta}(k)=-\frac{2a}{k^4}P^{(2)}-\frac{a}{2wk^4}P^{(0-s)}+(\cdots).
\end{equation}
Here $P^{(2)}$ and $P^{(0-s)}$ represent the Barnes-Rivers operators
\begin{eqnarray} \label{propto11}
P^{(2)}_{\mu\nu,\alpha\beta}&=&\frac{1}{2}(\theta_{\mu\alpha}\theta_{\nu\beta}+\theta_{\mu\beta}\theta_{\nu\alpha})-\frac{1}{3}\theta_{\mu\nu}\theta_{\alpha\beta},\\
P^{(0-s)}_{\mu\nu,\alpha\beta}&=&\frac{1}{3}\theta_{\mu\nu}\theta_{\alpha\beta},~~\theta_{\mu\nu}=\eta_{\mu\nu}-\frac{k_\mu k_\nu}{k^2},\label{propto12}
\end{eqnarray}
while $(\cdots)$ denotes the set of terms that are irrelevant to the spectrum of the theory. Note  that (\ref{propto1}) without $(\cdots)$ represents  a gauge-invariant part of the propagator. Also it  does not include  $1/k^2$ propagator appearing in the FDG.
The propagator (\ref{propto1}) carries  6 DOF: two massless tensors (4 DOF) $+$ two massless scalars (2 DOF) since the former propagator $1/k^4$ is the degenerate limit of the massless tensor  pole and the massive ghost pole (a massless tensor dipole) as well as the later one  is the degenerate limit of the massless scalar pole and the massive scalar pole (a massless dipole). The 6 DOF is compared to 8 DOF for the FDG.

By setting $k_0=0$, the spatial part
of the gauge-invariant propagator (\ref{propto1})  takes the form
\begin{eqnarray}
{\cal P}^{\rm SIG}_{\mu\nu,\kappa\lambda}(\mathbf{k})=&-&\frac{2a}{\mathbf{k}^4}\Big[\frac{1}{2}(\eta_{\mu\kappa}\eta_{\nu\lambda}+\eta_{\mu\lambda}\eta_{\nu\kappa})-\frac{1}{3}\eta_{\mu\nu}\eta_{\kappa\lambda}\Big] \label{propa1} \\
&-& \frac{a}{\mathbf{k}^4}\frac{ \eta_{\mu\nu}\eta_{\kappa\lambda}}{6w} \nonumber.
\end{eqnarray}
We note the relation between the classical potential sourced by a static  mass $M$ and propagator
\begin{equation}
V(r)=\frac{M}{4(2\pi)^3}\int d^3\mathbf{k} e^{i\mathbf{ k}\cdot\mathbf{ r}} {\cal P}_{00,00}(\mathbf{k}).
\end{equation}
Fourier-transforming
\begin{equation}\label{propa1z}
  {\cal P}^{\rm SIG}_{00,00}(\mathbf{k})=-\frac{4a}{3}\frac{1}{\mathbf{k}^4}-\frac{a}{6w}\frac{1}{\mathbf{k}^4}
\end{equation}
 leads to the classical linear potential   as
\begin{equation} \label{pot1}
V^{\rm SIG}(r)= \frac{Ma}{24\cdot\pi}\Big[1+\frac{1}{8w}\Big]r,
\end{equation}
where one uses  the relation~\cite{Alvarez-Gaume:2015rwa}
\begin{equation}
\int d^3\mathbf{k} \frac{ e^{i\mathbf{ k}\cdot\mathbf{ r}}}{\mathbf{k}^4}=\frac{2\pi}{r} r^2 \int^{\infty}_{-\infty}d(kr)\frac{\sin(kr)}{(kr)^3}=-\pi^2r.
\end{equation}
This classical potential $V^{\rm SIG}(r)$  is 0 (finite) at the origin of $r=0$ and the latter part of this potential was found in Ref.~\cite{Alvarez-Gaume:2015rwa}.
Also, it is desirable to discuss the Newtonian potential for the conformal gravity. Yoon~\cite{Yoon:2013rxa} criticized that Mannheim's conformal gravity~\cite{Mannheim:2005bfa} leads to negative linear potential,
 which is problematic from the point of view of fitting galaxy rotation curves. This requires positive linear potential.
However, Mannheim~\cite{Mannheim:2015gba} answered that Yoon had made an error in his analysis of the sign of linear coefficient.
In this work, from the first term in (\ref{pot1}), we read off  positive linear potential when choosing positive $a$.

On the other hand, adding  $2 R/\kappa^2$ to Eq.(\ref{action2}) leads to the FDG propagator~\cite{Myung:2017zsa}
\begin{equation} \label{propto2}
{\cal P}^{\rm FDG}_{\mu\nu,\alpha\beta}(k)=\Big[\frac{1}{k^2}-\frac{1}{k^2-m^2_2}\Big]P^{(2)}-\frac{1}{2}\Big[\frac{1}{k^2}-\frac{1}{k^2-m^2_0}\Big]P^{(0-s)}+(\cdots),
\end{equation}
where mass squared forms  are given by
\begin{equation}
m^2_2=\frac{2 a}{\kappa^2},~~m^2_0=-\frac{1}{\kappa^2}\frac{a}{w}.
\end{equation}
Here we require the non-tachyonic masses of $m^2_2>0$ and $~m^2_0>0$ which determines the signs of $a$ and $w$  such  that  $a>0$ and $w<0$.
This inequality implies  the positivity of the coefficient in front of $r$ in $V^{\rm SIG}$ (\ref{pot1}).
Then, we obtain  the famous classical potential
\begin{equation} \label{pot2}
V^{\rm FDG}(r)= -\frac{GM}{r}\Big[1-\frac{4}{3}e^{-m_2r}+\frac{1}{3}e^{-m_0r}\Big]
\end{equation}
when using
\begin{equation}
V^{\rm FDG}(r)=\frac{\kappa_4M}{(2\pi)^3}\int d^3\mathbf{k} e^{i\mathbf{ k}\cdot\mathbf{ r}} {\cal P}^{\rm FDG}_{00,00}(\mathbf{k}),
\end{equation}
where
\begin{equation}
 {\cal P}^{\rm FDG}_{00,00}(\mathbf{k})=\frac{1}{2}\Big[-\frac{1}{\mathbf{k}^2}+\frac{4}{3}\frac{1}{\mathbf{k}^2+m_2^2}-\frac{1}{3}\frac{1}{\mathbf{k}^2+m_0^2}\Big].
\end{equation}
Here we point out that in the limit of $r\to 0$, a massive ghost tensor contributes  $4/3(=1+1/3)$ to the potential
 and  a massive healthy scalar contributes $- 1/3$ to the potential.  The singularity cancellation occurs in the FDG.
This model shows that the theory without any kind of nonlocality could be free from the Newtonian singularity.
Interestingly, for $r\gg r_0$ where $r_0 =1/\min(m_2,m_0)$, we get the usual Newtonian potential of $V^{\rm FDG}\approx-GM/r$.
However, for $r\ll r_0$, one finds that
\begin{eqnarray}
V^{\rm FDG} &\approx& \frac{GM(m_0-4m_2)}{3}+\frac{GM(4m_2^2-m^2_0)r}{6}+\cdots\nonumber  \\
  &=&\frac{GM(m_0-4m_2)}{3} + \frac{Ma}{24\cdot\pi}\Big[1+\frac{1}{8w}\Big]r+\cdots.
\end{eqnarray}
Here we observe that the last term  is the same form as  (\ref{pot1}). Considering the condition of $a>0$ and $w<0$, the coefficient of $r$ is positive (negative) for $w<-0.125(w>-0.125)$.
It is worth noting that the classical potential $V^{\rm SIG}$ of the SIG is embedded as the linear term of the FDG when expanding $V^{\rm FDG}(r)$ around $r=0$.
This may suggest a deep renormalization connection between SIG and FDG~\cite{Stelle:1976gc,Julve:1978xn}.

\section{Renormalization of SIG}
We start by  mentioning that the three coupling constants $a,w,c$ are dimensionless in (\ref{action1}). Considering a propagator behaving as $1/k^4$,
the SIG has been shown to be renormalizable and asymptotically free  without including the Einstein-Hilbert term $2R/\kappa^2$.
Also, since  the static potential (\ref{pot1}) associated with $1/k^4$-propagator is proportional to distance $r$, this would be a confining theory.

Before we proceed, we have  to mention that the main purpose of \cite{Fradkin:1981iu} was the one-loop computation in the general theory including the Einstein-Hilbert  and cosmological terms,
and such a calculation was first performed in~\cite{Julve:1978xn}.

Let us remind  that the effect of $cGB$ term  is not relevant for one-loop renormalization, but it  may play a role in renormalization group equations.
However, the inclusion of $cGB$ made  a little difference in some coefficients of the beta-functions~\cite{deBerredoPeixoto:2004if,deBerredoPeixoto:2003pj}.

Hence, we  consider the renormalization group equations in terms of two coupling constants ($a,w$) which are relevant to determining the classical potential (\ref{pot1})~\cite{Einhorn:2014bka}.
In this case, the Weyl coupling $a$ may be identified with the loop-expansion parameter, whereas the Ricci  coupling $w$ will be seen to approach a UV fixed point.
We renormalize the couplings $a$ and $w$ by introducing  $\mu$ the renormalization parameter of dimensional regularization $(n=4-\epsilon)$.
With $(4\pi)^2 t=\ln[\mu/\mu_0]$, the relevant beta-functions take the forms
\begin{eqnarray}
\frac{da}{dt}&=&-\epsilon a -a^2\frac{\partial(aA_1)}{\partial a},\\
\frac{dw}{dt}&=&a\frac{\partial[a(B_1-wA_1)]}{\partial a}.
\end{eqnarray}
The one-loop divergences have provided the counterterms of $A_1$ and $B_1$ as~\cite{Avramidi:1985ki,deBerredoPeixoto:2003pj}
\begin{equation}
A_1=\frac{133}{10},~~ B_1=\frac{10w^2}{3}-5w+\frac{5}{12}.
\end{equation}
Considering $a$ as a loop-expansion parameter with tree approximation of order $1/a$,
$A_1$ and $B_1$ imply that  the one-loop divergences are independent of $a$.
In the limit of $\epsilon\to 0(n\to4)$, one finds the one-loop beta-functions
\begin{eqnarray} \label{rge1}
\frac{da}{dt}&=&-\frac{133}{10}a^2,\\
\frac{dw}{dt}&=&\Big[\frac{5}{12}-w\Big(5+\frac{133}{10}\Big)+\frac{10}{3}w^2\Big]a. \label{rge2}
\end{eqnarray}
The first equation (\ref{rge1}) is a negative beta-function and thus,  can be solved to be
\begin{equation}
a(t)=\frac{a_0}{1+\frac{133}{10}a_0 t},
\end{equation}
which shows that in the UV limit of $t\to \infty$, it manifests the asymptotic freedom for $a$.
That is, the Weyl coupling $a$ is always asymptotically free from solely gravity self-interactions.
 For the Ricci coupling $w$, we find  two roots of $0<w_1<w_2$ $(w_2=5.47,w_1=0.023)$ from the beta-function (\ref{rge2}).
The coupling $w$ approaches the UV fixed point $w_1$ for $w$ within the UV attraction region $0<w<w_2$.
A UV fixed point is located  at $w=w_1$  and so, the coupling $w$ is also asymptotically free. Also, an IR fixed point is found at $w=w_2$.
We note that the $w\to0$ limit  denotes a strong coupling limit where the one-loop analysis is not reliable~\cite{Holdom:2015kbf}.
Hence, the action (\ref{action2}) is asymptotically free for $a>0$ and $w>0$ and thus, it is  regarded as  a UV complete theory.

\section{Discussions}
First of all, we have found the linear classical potential $V^{\rm SIG}(r)$  (\ref{pot1}) from the SIG which  is zero (finite) at the origin of $r=0$.
The SIG provides another gravity,  showing that the higher-derivative gravity  without any kind of nonlocality  could be free from the Newtonian singularity.
This is because it had a massless tensor and scalar dipole propagator of $1/k^4$.
On the other hand, the SIG action (\ref{action1}) is renormalizable and  asymptotically  free for $a>0$ and $w>0$.
The above statements  show clearly that  the SIG  satisfies the conjecture which states that renormalzable higher-derivative gravity has  a finite classical potential at the origin.

Furthermore, we observe that  the inequality condition of $a>0$ and $w<-0.125$ is compatible with the positivity of coefficient in front of $r$ in $V^{\rm SIG}$, implying that the SIG is a confining theory.
In addition, it was known that the classical potential $V^{\rm SIG}$  is embedded as the linear term of the FDG when expanding $V^{\rm FDG}(r)$ around $r=0$.
This may suggest a deep connection between SIG  and FDG renormalizations.

Initially, we have included the Gauss-Bonnet (GB) term to investigate  its roles in both classical potential and one-loop renormalization in four dimensions.
Taking into account the GB term $cGB$, it did not contribute to the graviton propagator on the Minkowski spacetimes and thus, to  the classical potential.
On the renormalization  side, the effects of the Gauss-Bonnet term is not relevant for one-loop renormalization  and renormalization group equation in four dimensions~\cite{Avramidi:1985ki,deBerredoPeixoto:2004if}.
This implies that the GB term which does not correspond to any graviton interaction in four dimensions, plays no crucial role in both classical potential and one-loop renormalization in four dimensions.
That is, its role is trivial.

At this stage, we wish to discuss the unitarity issue of the SIG. As was shown in (\ref{propto1}), the former propagator of $1/k^4$ is considered as the degenerated limit of massless spin-2 pole and massive spin-2 ghost pole (massless tensor dipole) and the latter is the degenerated limit of massless spin-0 pole and massive spin-0 ghost pole (massless scalar dipole). This implies that the SIG has ghosts. It was known that there are two possible ways to avoid ghosts as followed by FDG : nonlocal FDG~\cite{Modesto:2011kw,Biswas:2011ar} or Lee-Wick gravity~\cite{Modesto:2015ozb}. However, the SIG has no the Einstein-Hilbert term, implying that  it gives rise to difficulty to remove the ghost-like states. Even though the SIG is renormalizable and asymptotically free, this theory is non-unitary.

Finally, the conformal gravity with $w=0$ in (\ref{action1}) provides the classical linear potential $V^{\rm CG}=[Ma/24\pi] r$,
while it is renormalizable without $R^2$ term~\cite{deBerredoPeixoto:2003pj}. For $a>0$, this gravity  is a confining theory as well as  a UV complete theory.
Hence, the conformal gravity  is regarded as  the simplest model which satisfies the conjecture.

\section*{Acknowledgement}
This work was supported by the National Research Foundation of Korea (NRF) grant funded by the Korea government (MOE)
 (No. NRF-2017R1A2B4002057).

\end{document}